\documentclass{article}
\usepackage{a4wide} 
\usepackage{amssymb,amsmath}
\usepackage{graphicx}
\newtheorem{theorem}{Theorem}

\begin{document}

\begin{center}

\begin{Large}\begin{bf}
A Comment on the Relation between\\[1ex] Diffraction and Entropy\vspace{3ex}
\end{bf}\end{Large}

\begin{large}
Michael Baake$^{1}$ and Uwe Grimm$^{2}$\vspace{3ex}
\end{large}

\begin{small}
$^{1}$Fakult\"{a}t f\"{u}r Mathematik, Universit\"{a}t Bielefeld,\\
       Postfach 100131, 33501 Bielefeld, Germany\\[0.5ex]
$^{2}$Department of Mathematics and Statistics, The Open University,\\
       Walton Hall, Milton Keynes MK7 6AA, United Kingdom\vspace{1ex}
\end{small}
\end{center}

\begin{quote}Diffraction methods are used to detect atomic order in
  solids. While uniquely ergodic systems with pure point diffraction
  have zero entropy, the relation between diffraction and entropy is
  not as straightforward in general.  In particular, there exist
  families of homometric systems, which are systems sharing the same
  diffraction, with varying entropy. We summarise the present state of
  understanding by several characteristic examples.
\end{quote}

\section{Introduction}

Quantifying order or complexity in systems is a difficult task, as
there are no universal measures of order or complexity. In the realm of
solid state physics, atomic order is usually probed by diffraction
experiments \cite{C}. A pure point diffraction measure, which means a
diffraction pattern comprising only Bragg peaks and no continuous
component, is an indicator of atomic order. Crystalline solids are
paradigms of pure point diffractive systems, and diffraction
experiments are used to determine the symmetry and atomic structure of
the crystal. More generally, perfect quasicrystals are
also pure point diffractive, even though the diffraction peaks are
located on a Fourier module that is dense in space. Nevertheless, at
any experimental resolution, only finitely many diffraction peaks are
resolved in any finite region of space, so the diffraction pattern
appears discrete in practice. This is one commonly accepted model for
real world quasicrystals \cite{SBGC84}.

Another measure of order versus complexity is the configurational
entropy of a system \cite{CFS}. In an ideal crystal, atomic positions
are determined by those within a fundamental domain of the underlying
lattice, and hence the configurational entropy is zero. This is also
true for perfect quasicrystals, and pure point diffraction is indeed
related to zero entropy, as shown in \cite{BLR} for a large class of
systems.

It is interesting to explore what happens with the relation between
diffraction and entropy if one leaves the pure point regime; see
\cite{HB00} and references therein for examples. With increasing
experimental resolution, continuous diffraction intensities are
becoming more accessible (see \cite{Withers} for a recent exposition),
and it is important to understand their origin and the implications on
the structure of the material under investigation. This is a first
step in tackling the inverse problem of diffraction in the more
general setting of mixed diffraction spectra.

In this short note, we aim to highlight the scope of the inverse
problem, by presenting a variety of examples with continuous
diffraction. These reveal that the picture is indeed rather complex.
In particular, we recall a family of homometric (or isospectral)
structures which cover a full available range of configurational
entropy, from a (fully deterministic) case with zero entropy to a
completely random case. In higher dimensions, a variety of
possibilities exist, including lower rank entropy.

Most of the results below have appeared in original papers, but not in
one source. We use this note to review some of the more recent
attempts, and try to put them in a more systematic frame.  After a
brief recapitulation of mathematical diffraction theory, we proceed
along this path mainly by way of characteristic examples.

\section{Diffraction of weighted Dirac combs}

Mathematical diffraction theory was pioneered by Hof in \cite{Hof,Hof2},
and should be considered as the rigorous mathematical counterpart
of kinematic diffraction; compare \cite{C} for background.
For simplicity, we concentrate on the diffraction of weighted Dirac
combs \cite{Crelle,BG11}.  A weighted Dirac comb of a general point
set $S\subset\mathbb{R}^{d}$ is formally spelled out as
\[
   \omega \, = \, \sum_{x\in S} w(x)\,\delta_{x}
   \, = \, w\,\delta^{}_{S} \, ,
\]
where $\delta_{x}$ is the normalised point (or Dirac) measure at $x$,
and $w(x)$ is a weight function (which may be complex). Here,
$\delta^{}_{S}:= \sum_{x\in S} \delta_{x}$ is the Dirac comb of $S$.
We assume that the set $S$ and the weight function $w$ are such that
the corresponding weighted Dirac comb $\omega$ is a translation
bounded measure, and that its natural \emph{autocorrelation measure}
\begin{equation}\label{eq:auto}
   \gamma \, =  \, \gamma^{}_{\omega} 
                  = \, \omega \circledast \widetilde{\omega} 
                 \, := \lim_{R\to\infty}
                \frac{\;\omega|^{}_{R} * \widetilde{\omega|^{}_{R}}\;}
                   {\mathrm{vol} (B_R)} ,
\end{equation}
exists. Here, $B_{R}$ denotes the open ball of radius $R$ around $0\in
\mathbb{R}^{d}$ and $\omega|^{}_{R}$ the restriction of $\omega$ to
$B_{R}$.  For a measure $\mu$, its `flipped-over' version
$\widetilde{\mu}$ is defined via $\widetilde{\mu}(g) =
\overline{\mu(\widetilde{g})}$ for $g\in
C_{\mathsf{c}}(\mathbb{R}^{d})$, where
$\widetilde{g}(x)=\overline{g(-x)}$.  The volume-averaged (or
Eberlein) convolution $\circledast$ is needed because $\omega$ itself
generally is an unbounded measure, so the direct convolution is not
defined. For instance, if $\lambda$ denotes the standard Lebesgue
measure (for volume), $\lambda\ast\lambda$ is not defined, while
$\lambda\circledast\lambda=\lambda$.  Note that different measures
$\omega$ can share the same autocorrelation $\gamma$. This phenomenon
is called \emph{homometry}, and we shall see explicit examples later
on.

The autocorrelation measure $\gamma$ is \emph{positive definite} (or
of positive type) by construction, which means $\gamma(g *
\widetilde{g}\, )\ge 0$ for all $g\in
C_{\mathsf{c}}(\mathbb{R}^{d})$. As a consequence, its Fourier
transform $\widehat{\gamma}$ exists \cite{BF} and is a translation
bounded, positive measure, called the \emph{diffraction measure} of
$\omega$. It describes the outcome of kinematic diffraction of
$\omega$ by quantifying how much scattering intensity reaches a given
volume in $d$-space; see \cite{Hof,BG11,TAO} for more details.

Relative to Lebesgue measure $\lambda$, we have the unique
splitting
\[
   \widehat{\gamma} \; = \; 
            \widehat{\gamma}^{}_{\mathsf{pp} } +
            \widehat{\gamma}^{}_{\mathsf{sc} } +
            \widehat{\gamma}^{}_{\mathsf{ac} }
\]
of $\widehat{\gamma}$ into its pure point part (the Bragg peaks, of
which there are at most countably many), its absolutely continuous
part (the diffuse scattering with locally integrable density relative
to $\lambda$) and its singular continuous part (which is whatever
remains). The last contribution, if present, is described by a measure
that gives no weight to single points, but is still concentrated to an
uncountable set of zero Lebesgue measure. Examples of such measures
are provided by the Thue-Morse system and its generalisations; see
\cite{BG08,BGG12} and references therein.

\section{Bernoullisation}

The classic coin tossing process leads to the Dirac comb $\omega =
\sum_{n\in\mathbb{Z}} X(n) \, \delta_{n}$, where the
$(X(n))^{}_{n\in\mathbb{Z}}$ form an i.i.d.\ family of random
variables, each taking values $1$ and $-1$ with probabilities $p$ and
$1-p$, respectively.  By an application of the strong law of large
numbers (SLLN), almost every realisation has the autocorrelation
measure
\[
    \gamma \, = \, (2p-1)^{2} \, \delta^{}_{\mathbb{Z}}
    + 4 p (1-p) \, \delta^{}_{0} \, ,
\]
and hence (via Fourier transform) the diffraction measure
\[
    \widehat{\gamma} \, = \, (2p-1)^{2} \, \delta^{}_{\mathbb{Z}}
    + 4 p (1-p) \, \lambda \, .
\]
Here, we have used the classic Poisson summation formula
$\widehat{\delta^{}_{\mathbb{Z}}} = \delta^{}_{\mathbb{Z}}$; compare
\cite{BG11} and references therein for a formulation in the
diffraction context.  When $p=\frac{1}{2}$, the diffraction boils down
to $\widehat{\gamma} = \lambda$. Here, the point part is extinct
because the average scattering strength vanishes. For proofs, we refer
to \cite{BM98,BBM}.

The Bernoulli chain has (metric) entropy $H(p) = - p\log (p) -
(1\!-\!p) \log (1\!-\!p)$, which is maximal for $p=\frac{1}{2}$, with
$H(\frac{1}{2})=\log (2)$. It vanishes for the deterministic limit
cases $p\in\{0,1\}$. For the latter, we have
$\omega=\mp\delta^{}_{\mathbb{Z}}$, and consequently obtain the
diffraction measure $\widehat{\gamma} = \delta^{}_{\mathbb{Z}}$, again
via the Poisson summation formula.

In contrast, the (binary) Rudin-Shapiro chain is a deterministic
system, with polynomial complexity function and thus zero entropy. The
corresponding sequence of weights $(w(n))^{}_{n\in\mathbb{Z}}$ with
$w(n)\in\{\pm 1\}$ can be defined recursively by the initial
conditions $w(-1)=-1$, $w(0)=1$, together with
\begin{equation}\label{eq:rs}
   w(4n+\ell)=
    \begin{cases} w(n),  & \mbox{for $\,\ell\in\{0,1\}$,} \\
          (-1)^{n+\ell}\,w(n), & \mbox{for $\,\ell\in\{2,3\}$,}
     \end{cases}
\end{equation}
which determines $w(n)$ for all $n\in\mathbb{Z}$.  Despite its
deterministic nature, the autocorrelation measure is simply given by
$\gamma^{}_{\mathrm{RS}} = \delta^{}_{0}$, so that
$\widehat{\gamma^{}_{\mathrm{RS}}} = \lambda$; see \cite{BG09,BG11} 
for further details and a simple proof. Alternatively, the result
also follows from the exposition in \cite{Q,PF}.

Now, the theory of random variables allows for an interpolation
between the two cases as follows. Let us 
consider the random Dirac comb
\begin{equation}\label{eq:randrs}
   \omega_{p} \, = \sum_{n\in\mathbb{Z}} w(n)\, X(n)\,\delta_{n}\, ,
\end{equation}
where $(X(n))^{}_{n\in\mathbb{Z}}$ is, as above, an i.i.d.\ family of
random variables with values in $\{\pm 1\}$ and probabilities $p$ and
$1-p$. This `Bernoullisation' of the Rudin-Shapiro comb can be viewed
as a model of second thoughts, where the sign of the weight at
position $n$ is changed with probability $1-p$. By a (slightly more
complicated) application of the SLLN, it can be shown \cite{BG09} that
the autocorrelation $\gamma$ of the Dirac comb $\omega$ is almost
surely given by
\[
   \gamma_{p}  \, = \, (2p-1)^{2}\,\gamma^{}_{\mathrm{RS}} + 
   4 p (1-p)\, \delta^{}_{0}
   \, = \, \delta^{}_{0}\, ,
\]
irrespective of the value of the parameter $p\in [0,1]$. This
establishes the following result; see \cite{BG09,BG11} for details.

\begin{theorem}
   The family of random Dirac combs\/ $\omega_{p}$ of
   Eq.~\eqref{eq:randrs} with\/ $p\in [0,1]$ are (almost surely)
   homometric (isospectral), with absolutely continuous diffraction
   measure\/ $\widehat{\gamma_{p}} = \widehat{\gamma^{}_{\mathrm{RS}}}
   = \lambda$, irrespective of the value of\/ $p$.
\end{theorem}

This result shows that diffraction can be insensitive to entropy,
because the family of Dirac combs $\omega_{p}$ of
Eq.~\eqref{eq:randrs} continuously interpolates between the
deterministic Rudin-Shapiro case with zero entropy and the completely
random Bernoulli chain with maximal entropy $\log(2)$. Clearly, this
example can be generalised to other sequences, and (by taking
products) to higher dimensions.

\section{Close-packed dimers}

Another instructive example in one dimension was recently suggested by
van Enter \cite{BE}. Partition $\mathbb{Z}$ into a close-packed
arrangement of `dimers' (pairs of neighbours), without gaps or
overlaps. Clearly, there are two possibilities to do so. Next,
decorate each pair randomly with either $(1,-1)$ or $(-1,1)$, with
equal probability. The set of all sequences defined in this way is
given by 
\[
   \mathbb{X} \, = \, \bigl\{ w \in \{ \pm 1 \} ^{\mathbb{Z}} \mid
   M(w) \subset 2 \mathbb{Z} \,\text{ or }
   M(w) \subset 2 \mathbb{Z} + 1 \bigr\} \, ,
\]
where $M(w) := \{ n\in\mathbb{Z} \mid w(n) = w(n+1) \}$. Note that
$M(w)$ is empty precisely for the two periodic sequences $w(n)=\pm
(-1)^n$.

Considering the corresponding signed Dirac comb on $\mathbb{Z}$ with
weights $w(n)\in\{\pm 1 \}$, it can be shown that
its autocorrelation almost surely exists and is given by \cite{BE}
\[
   \gamma\, =\, \delta^{}_{0} - \frac{1}{2} 
              (\delta^{}_{1} + \delta^{}_{-1})\, .
\]
The corresponding diffraction measure is then
\begin{equation}\label{eq:dms}
   \widehat{\gamma} \,=\,
   \bigl( 1 - \cos(2 \pi k) \bigr) \lambda\, ,
\end{equation}
which is again a purely absolutely continuous diffraction
measure. Here, the continuous density relative to $\lambda$ is written
as a function of $k$.

On first sight, the system looks disordered, with entropy of
$\frac{1}{2} \log (2)$. This seems (qualitatively) reflected by the
diffraction. However, the system also defines a dynamical system under
the action of $\mathbb{Z}$, as generated by the shift $S \! : \,
\mathbb{X} \longrightarrow \mathbb{X}$, with $(Sw) (n) := w(n+1)$. As
such, it has a dynamical spectrum that does contain a pure point part,
with eigenvalues $0$ and $\frac{1}{2}$; we refer to \cite{Q} for
general background on this concept, and to \cite{BE} for the actual
calculation of the eigenfunctions. The extension to a dynamical system
under the general translation action of $\mathbb{R}$ is a standard
procedure known as suspension; see \cite[Ch.~11.1]{CFS} for an
introduction, where the suspension is called a special flow.

This finding suggests that some degree of order must be present that
is neither visible from the entropy calculation nor from the
diffraction measure alone. Indeed, one can define a factor of the
system by a continuous mapping $\phi\!:\, \mathbb{X}\longrightarrow \{\pm
1\}^{\mathbb{Z}}$ defined by $(\phi w)(n) = -w(n)w(n+1)$. It maps
$\mathbb{X}$ globally 2:1 onto 
\[
  \mathbb{Y}=\phi(\mathbb{X})=\bigl\{v\in\{\pm 1\}^{\mathbb{Z}}\mid
  \mbox{$v(n)=1$ for all $n\in2\mathbb{Z}$ or all $n\in 2\mathbb{Z}+1$}\bigr\}.
\]
The autocorrelation and diffraction measure of the signed Dirac comb
$v\delta^{}_{\mathbb{Z}}$ for an element $v\in\mathbb{Y}$ are almost
surely given by
\[
    \gamma \, = \, \frac{1}{2}\delta^{}_{0} + 
    \frac{1}{2}\delta^{}_{2\mathbb{Z}}
    \quad\text{and}\quad
    \widehat{\gamma}\, =\, \frac{1}{2}\lambda + 
    \frac{1}{4}\delta^{}_{\mathbb{Z}/2}\, .
\]
The diffraction of the factor system $\mathbb{Y}$ uncovers the
`hidden' pure point part of the dynamical spectrum, which was absent
in the purely absolutely continuous diffraction of the signed Dirac
comb $w\delta^{}_{\mathbb{Z}}$ with $w\in\mathbb{X}$. In summary, we
have the following situation \cite{BE}.

\begin{theorem}
  The diffraction measure of the close-packed dimer system\/
  $\mathbb{X}$ with balanced weights is purely absolutely continuous
  and given by Eq.~\eqref{eq:dms}.

  The dynamical spectrum of the close-packed dimer system\/
  $\mathbb{X}$ under the translation action of\/ $\mathbb{R}$ contains
  the pure point part\/ $\mathbb{Z}/2$ together with a countable
  Lebesgue spectrum.

  The non-trivial part\/ $\mathbb{Z} + \frac{1}{2}$ of the dynamical
  point spectrum is not reflected by the diffraction spectrum of\/
  $\mathbb{X}$, but can be recovered via the diffraction spectrum of a
  suitable factor, such as\/ $\mathbb{Y}$.
\end{theorem}
A similar observation can be made for the (generalised) Thue-Morse
system; see \cite{EM,BGG12}.

\section{Ledrappier's model}

For a long time, people had expected that higher dimensions
are perhaps more difficult, but not substantially different.
This turned out to be a false premise though, as can be
seen from the now classic monograph \cite{Sch}.

In our present context, we pick one characteristic example, the system
due to Ledrappier \cite{L}, to show one new phenomenon. Here, we consider a
specific subset of the full shift space $\{\pm 1\}^{\mathbb{Z}^{2}}$,
defined by
\begin{equation}\label{eq:def-L}
  \mathbb{X}_{\mathrm{L}} \, =\,
  \bigl\{ w \in \{\pm 1\}^{\mathbb{Z}^{2}} \! \mid
     w(x)\, w(x+e^{}_{1})\, w(x+e^{}_{2}) = 1 \,
     \mbox{ for all }\, x \in \mathbb{Z}^{2} \bigr\},
\end{equation}
where $e^{}_{1}$ and $e^{}_{2}$ denote the standard Euclidean basis
vectors in the plane. On top of being a closed subshift,
$\mathbb{X}_{\mathrm{L}}$ is also an Abelian group (here written
multiplicatively), which then comes with a unique, normalised Haar
measure. The latter is also shift-invariant, and the most natural
measure to be considered in our context.

The system is interesting because the number of patches of a given
radius (up to translations) grows exponentially in the radius rather
than in the area of the patch. This phenomenon is called \emph{entropy
  of rank $1$}, and indicates a new class of systems in higher
dimensions. More precisely, along any lattice direction of
$\mathbb{Z}^{2}$, the linear subsystems essentially behave like
one-dimensional Bernoulli chains. It is thus not too surprising that
the diffraction measure satisfies the following theorem, though its
proof \cite{BW10} has to take care of the special directions connected
with the defining relations of $\mathbb{X}_{\mathrm{L}}$.

\begin{theorem}
  If $w$ is an element of the Ledrappier subshift\/ $\mathbb{X}_{\mathrm{L}}$
  of Eq.~\eqref{eq:def-L}, the corresponding weighted Dirac comb
  $w \delta^{}_{\mathbb{Z}^2}$ has diffraction measure $\lambda$, which
  holds almost surely relative to the Haar measure of\/
  $\mathbb{X}_{\mathrm{L}}$.
\end{theorem}

So, the Ledrappier system is homometric to the full
$\mathbb{Z}^{2}$-shift, which means that an element of either system
almost surely has diffraction measure $\lambda$. As mentioned before,
via a suitable product of two Rudin-Shapiro chains, also a
deterministic system with diffraction $\lambda$ exists. This clearly
demonstrates the insensitivity of pair correlations to the (entropic)
type of order or disorder in the underlying system.

Although correlation functions of higher order can resolve the
situation in this case, one can consider other dynamical systems
(such as the ($\times 2,\times 3$)-shift \cite{BW10}) that share
almost all correlation functions with the Bernoulli shift on
$[0,1]^{\mathbb{Z}^{2}}$. This is a clear indication that our present
understanding of `order' is incomplete, and that we still lack
a good set of tools for the detection of order.

\section{Meyer sets with entropy}

Meyer sets in Euclidean space are point sets $\varLambda \subset
\mathbb{R}^{d}$ that are relatively dense in such a way that
$\varLambda - \varLambda$ is still uniformly discrete. This innocently
looking condition has deep consequences \cite{Meyer,Bob,Jeff}.  In
particular, it is reasonable to consider Meyer sets as natural
generalisations of lattices. They comprise perfect quasicrystals (as
those obtained from the projection method), but are general enough to
accommodate entropy as well.

As a simple example, start from the set $2 \mathbb{Z}$ and add any
subset of $2 \mathbb{Z} + 1$ to it, for instance a random selection of
the latter. This is a Meyer set (it contains $2 \mathbb{Z}$, so that
it is relatively dense, while the Minkowski difference is a subset of
$\mathbb{Z}$, hence uniformly discrete).  Nevertheless, such a set has
entropy.  More generally, even though deterministic Meyer sets are the
ones that have been studied in most detail so far, `most' Meyer sets
will have entropy, but still possess a high degree of intrinsic order.
This is manifest from the following observation of Strungaru
\cite{S05}.

\begin{theorem}
  Let $S\subset \mathbb{R}^{d}$ be a Meyer set and $\omega :=
  \delta^{}_{S}$ the corresponding Dirac comb. If $\gamma$ is any
  autocorrelation of $\omega$, its Fourier transform
  $\widehat{\gamma}$ comprises a non-trivial pure point part. In
  particular, for any $\varepsilon > 0$, the set $\{ k \in 
  \mathbb{R}^{d} \mid \widehat{\gamma} (\{ k \}) \ge (1-\varepsilon)
  \, \widehat{\gamma} (\{0\}) \}$ is relatively dense.
\end{theorem}

In this sense, long-range order in Meyer sets leaves a remarkable
fingerprint. Considering subsets of a lattice, even without
demanding their relative denseness, a related result was also
proved in \cite{B02}. Let us take a closer look by means of a
famous example from number theory.

\section{Visible lattice points}

The visible (or primitive) points of the square lattice are defined as
\[
   \mathcal{V} \, = \, \{ (m,n) \in \mathbb{Z}^{2} \mid
   \gcd(m,n) = 1 \} \, .
\]
$\mathcal{V}$ is clearly uniformly discrete, but contains holes of
arbitrary size (as a consequence of the Chinese remainder theorem; see
\cite{BMP} for details). Consequently, $\mathcal{V}$ is neither a
Meyer nor a Delone set. Nevertheless, the set $\mathcal{V}$ has a
well-defined density ($6/\pi^2$), and positive topological entropy (of
the same value, if using the logarithm to base $2$). Moreover, one
also has the following result.

\begin{theorem}
  The Dirac comb $\delta^{}_{\mathcal{V}}$ has a pure point
  diffraction measure.
\end{theorem}

The proof of this claim in \cite{BMP} is constructive and also gives a
closed (and computable) formula for the diffraction measure.  In view
of \cite{BLR}, it is somewhat astonishing that pure point diffraction
and positive entropy go together like this. However, in a recent paper
by Huck and Pleasants \cite{HP}, it is shown that the natural
\emph{metric} entropy of $\mathcal{V}$ vanishes. The term `natural'
refers to the use of a nested sequence of growing discs as averaging
sequence; see \cite[Appendix]{BMP} for details.  The proof is again
constructive, and explains the mechanism:\ The frequencies of
arbitrary patches exist (though not uniformly so), which defines a
natural invariant measure via suitable cylinder sets. Now, a small set
of patches have large frequencies, while the majority sports small or
tiny frequencies -- and together this suffices to give metric entropy
$0$ (relative to this measure). The main point here is that the
frequencies (for the measure) and the pair correlations (for the
autocorrelation, and hence for the diffraction) are determined by
means of the \emph{same} averaging sequence, which clearly is the
relevant pairing.

Note that other invariant measures exist (for instance via different
averaging sequences), including examples with positive entropy.
It is not known what the matching diffraction measure would be,
but it is expected that they will show continuous components. A
careful analysis of all invariant measures for this example seems
an interesting open problem.

\section{Concluding remarks}

The examples above highlight different aspects of the quantification
of order in terms of entropy and diffraction. While pure point
diffractivity of uniquely ergodic systems \cite{BLR} implies zero
entropy, the general situation is complex, and there is no
straightforward relation between entropy and diffraction; in fact, as
the Bernoullisation example shows, diffraction can be completely
insensitive to the (entropic) disorder of a system.

In the example of the closed-packed dimers, we referred to the dynamical
spectrum (under the translation action). As this example 
together with the earlier observation in \cite{EM} shows, the
diffraction and dynamical spectra are, in general, not the same, and
can even have contributions of different spectral type. In the pure
point case, the notions are equivalent (in the sense that the
dynamical spectrum is pure point if and only if the diffraction
spectrum is pure point \cite{BL}), but in general the dynamical spectrum
contains additional information. It has been conjectured that the
latter should correspond to the diffraction spectra of the system and
all its factors.

Clearly, our understanding of 'order' is far from complete, and more
work is required to arrive at a clearer picture of what 'order' means,
and how to quantify it. Studying examples of the type discussed above
is a first step in this direction, and a general frame is explained
in \cite{Gou,BBM}. By mapping the range of possibilities, one gradually
obtains a better understanding of the plethora of manifestations of
order. This seems necessary in view of the hard inverse problem
for systems with diffuse scattering.

\begin{small}

\end{small}

\end{document}